\documentstyle[preprint,aps]{revtex}
\renewcommand{\baselinestretch}{1.1}
\draft
\begin{document}

\title{Sensitivity of tensor analyzing power in the process
$d+p\rightarrow
d+X$ to the longitudinal isoscalar form factor of the Roper resonance
electroexcitation}.
\author{E. Tomasi-Gustafsson \footnote{ E-mail:etomasi@cea.fr} and M. P.
Rekalo
\footnote{ Permanent address:
\it National Science Center KFTI, 310108 Kharkov, Ukraine}}
\address{\it DSM-CEA/IN2P3-CNRS, Laboratoire National Saturne and \\
DAPNIA/SPhN, C.E.A./Saclay,  91191 Gif-sur-Yvette Cedex, France}
\author{R. Bijker}
\address{Instituto de Ciencias Nucleares, U.N.A.M., A.P. 70-543, 04510 M\'exico
D.F., M\'exico}
\author{A. Leviatan}
\address{Racah Institute of Physics, The Hebrew University, Jerusalem 91904,
Israel}
\author{F. Iachello}
\address{Center for Theoretical Physics, Sloane Laboratory, Yale University, 
New Haven, CT 06520-8120, U.S.A.}
\maketitle

\begin{abstract}
The tensor analyzing power of the process $d + p \to d + X$, for forward
deuteron scattering in the momentum interval 3.7 to 9 GeV/c, is studied in
the framework of $\omega$ exchange in an algebraic collective model for the
electroexcitation of nucleon resonances. We point out a special sensitivity of
the tensor analyzing power to the
isoscalar longitudinal form factor of the Roper resonance excitation. The 
main argument is that the $S_{11}(1535)$, $D_{13}(1520)$ and
$S_{11}(1650)$ resonances have only isovector longitudinal form factors. It
is the longitudinal form factor of the Roper excitation, which plays an
important
role in the
$t-$dependence of the tensor analyzing power. We discuss possible evidence of
swelling of
hadrons with increasing excitation energy.
\end{abstract}

\section{Introduction}

In a previous paper \cite{Re96} it was shown that the polarization
observables in inclusive scattering of high energy deuterons by protons
at zero scattering angle, are sensitive to the ratio
$$
r=\sigma_L/\sigma_T,
$$
 where $\sigma_L$ and $\sigma_T$ are the cross
sections of
absorption of virtual isoscalar photons with longitudinal and transversal
polarizations by nucleons. In the framework of the
$\omega-$exchange mechanism for the considered reaction, it was found that
this sensitivity is especially large in the region of $N^*$-excitations
with masses $1.4-1.6$ GeV, for $r \le 0.5$.
This interval is especially interesting when compared with the data
obtained by
inclusive $eN-$scattering. This sensitivity is due, in particular, to the
properties of the
deuteron electromagnetic form factors. In \cite{Re96} a simplified
assumption was used, namely, the
ratio $r$ was taken to be a free parameter independent of the four momentum
transfer square,
$t=-q^2$. The  value that best fitted the data was $r=0.1$.
It is then interesting to compare this value with that of a realistic
model for the nucleon resonances electroexcitation.
In this work we present the results of the analysis of
polarization phenomena in $d+p\rightarrow d+X$, using the predictions
of the model \cite{Bi94,Bi96,Bi97} for
the $t$-dependence of the isoscalar form factors of electromagnetic
$N\rightarrow N^*$transitions. We show that hadronic probes of nucleon
structure, in particular, using polarized particles, may
give interesting and important information concerning form factors of
$N^*$-excitations. The selectivity of reactions such as $p(d,d')X$
or $p(\alpha,\alpha ')X$ to the isoscalar part of the
$N^*$-electroexcitation makes these processes complementary to
electron-nucleon inelastic scattering, for the study of the $N^*$-structure.
Note that in the framework of the $\omega$-exchange model \cite{Re96},
all polarization phenomena for $d+p\rightarrow d+X$ can be predicted
without any free parameters, using only existing information about
the deuteron electromagnetic form factors and about the ratio $r$.

This paper is organized as follows. In section II we give the $t$-dependence of
the tensor analyzing power
$T_{20}$ and $r$ in the  model [1]. Formulas for transverse and longitudinal
amplitudes of the algebraic collective model [2-4] are recalled in section
III. Results and discussions are presented in section IV and the conclusions
are drawn in section V.

\section{\boldmath{$\lowercase{t}$-dependence of $T_{20}$ and 
$\lowercase{r}$ }}

We will analyze here the polarization phenomena in the process $d+p\rightarrow
d+X$, for
forward deuteron scattering, in the framework of the model \cite{Re96} based on
$\omega-$exchange (Fig. 1). The $\omega$-meson is preferred, among the 
isoscalar mesons as
$\sigma$ or $\eta$, for several reasons. The $\omega NN-$ coupling is large; 
the $\omega$-meson,
being a spin 1 particle,  can induce strong polarization effects and an
energy-independent
cross section. When considered as an $isoscalar~photon$, then the cross
sections and the
polarization observables can be calculated from the known electromagnetic
properties of the deuteron and $N^*$, through
the vector dominance model.
These special properties of the $\omega-$exchange mechanism allow an
experimental test of the
validity of this model, similar to the Rosenbluth test of the one-photon
mechanism, in case of
elastic and inelastic electron-hadron scattering. The details of the model are
described in
\cite{Re96}. We will recall here only  the final expressions, necessary for the
present
analysis.

The tensor analyzing power in  $d+p\rightarrow d+X$, $T_{20}$, can be written 
in terms of the
electromagnetic form factors as:
\begin{equation}
T_{20}=-\sqrt{2}\frac{V_1^2+(2V_0V_2+V_2^2)r}{4V_1^2
+(3V_0^2+V_2^2+2V_0V_2)r},
\label{li1}
\end{equation}
where  $V_0(t)$, $V_1(t)$ and $V_2(t)$ are  related to the standard
electromagnetic deuteron
form factors: $G_c(t)$ (electric), $G_m(t)$ (magnetic) and $G_q(t)$ 
(quadrupole) by:
$$V_0=\sqrt{1+\tau}\left (G_c-\frac{2}{3}\tau G_q\right
),~~V_1=\sqrt{\tau}
G_m,~~V_2=\frac{ \tau} {\sqrt{1+\tau}}\left [-G_c+2\left (1-\frac{1}{3}\tau
\right )
G_q\right ],$$
and $\tau=-{t}/{4M_d^2}$, where $M_d$ is the deuteron mass.
The ratio $r$ characterizes the relative role of longitudinal and transversal
isoscalar
excitations in the transition $\omega+N\rightarrow X$. In case of the Roper
excitation we can
write:
\begin{equation}
r_R(t)=\frac{|A_{\ell}^p+A_{\ell}^n|^2}{|A_{1/2}^p+A_{1/2}^n|^2} ~,
\end{equation}
where $A_{\ell}^N~(A_{1/2}^N)$ is the longitudinal (transversal)
form factor of the $P_{11}(1440)$-excitation on proton ($N=p$) or neutron
($N=n$) targets. This formula can be
generalized to the excitation of any nucleon resonance $N^*$
as follows:
\begin{equation}
r_{N^*}(t)=\frac{|A_{\ell}^p+A_{\ell}^n|^2}
{|A_{1/2}^p+A_{1/2}^n|^2+|A_{3/2}^p+A_{3/2}^n|^2}\equiv
\sigma_L(t)/\sigma_T(t),
\end{equation}
where $A_{1/2}^{N}$ and $A_{3/2}^{N}$ are the two possible transversal
form factors, corresponding to
total $\gamma^*+N-$ helicity equal to $1/2$ and $3/2$ respectively.

In case of overlapping resonances, taking into account the finite values of the
resonance
widths, Eq. (3) can be generalized to
\begin{equation}
r\rightarrow
r(t,W)=\frac{\sum_i\sigma_{L,i}(t)B_i(W)C_i}{\sum_i\sigma_{T,i}(t)B_i(W)C_i},
\end{equation}
where $B_i(W)$ is a Breit-Wigner function for the $i-th~N^*$-resonance with a
definite
normalization:
\begin{equation}
C_i^{-1}=\int_{M+M_\pi}^\infty dW B_i(W),
\end{equation}
$M$ is the nucleon mass, $M_\pi$ is the pion mass and $W$ is the effective
invariant mass of
the $X-$system in $d+p\rightarrow d+X$ (i.e. the mass of the resonance).
Let us mention that for forward deuteron scattering in
$d+p\rightarrow d+X$ the variables $t$
and $W$ are not independent, for a fixed energy of the incoming deuteron
there is a definite correspondance between $t$ and $W$ \cite{Re96}.
The following observations, based on Eq. (1), can be made.
All information about the $\omega N N^*$-vertex is contained in the
function $r$ only. $T_{20}$ is especially sensitive to the small value
of $r(t,W)$ in the interval $0 \leq r\leq 0.5$. A zero value of $r$
results in a $t-$independent value for $T_{20}$, namely $T_{20}=-1/\sqrt{8}$,
for any value of
the deuteron electromagnetic form factors. The position of the points at
$q\simeq 1.85$~fm$^{-1}$  and  $q\simeq
5$~fm$^{-1}$, (with $q=\sqrt{-t}$), at which all theoretical curves
for different $r$ intersect,
is determined by the models used for the deuteron form factors.
In Fig. 2 we show the $t-$dependence of $r$, calculated on the basis of
Eq. (3), for two different cases: (i) considering the contribution of only
the Roper resonance (dotted line), (ii) considering the
sum of the contributions of the following resonances
\begin{equation}
P_{11}(1440),\;\; S_{11}(1535),\;\; D_{13}(1520), \;\; S_{11}(1650) ~,
\end{equation}
which are overlapping in this mass region (solid line). The form factors of the
$N^*$-excitations
on proton and neutron targets were derived using an algebraic
collective model \cite{Bi94,Bi96}, some details of which will be discussed
in the next section. This will then allow to understand the behaviour of the
ratio $r$ as shown in Fig. 2.

\section {Algebraic collective model}

\noindent
The formulas for transverse and longitudinal helicity
amplitudes used in the present work, are based on the collective string
model of baryons \cite{Bi94,Bi96,Bi97}, assuming $SU_{sf}(6)$ symmetry. In
this model the nucleon resonances are interpreted in terms of rotations and
vibrations of a Y-shaped string configuration with a prescribed distribution
of charges and magnetization. The underlying algebraic structure of the
model enables a derivation of closed expressions
for masses, electromagnetic and strong couplings of baryon resonances.
Electromagnetic transverse and longitudinal form factors on proton
and neutron targets for nucleon resonances below 2 GeV
are shown in Tables I-II as a function of
the photon momentum $k$.  The parameters relevant to these observables
are the constituent mass, $m=0.336$ GeV,
magnetic moment, $\mu=0.127$ GeV$^{-1}$ (g-factor $g$=1), and a scale
parameter of the distribution, $a=0.232$ fm. The state of the resonance is
recalled in Table II only.

As seen from Table II, of the four resonances mentioned in (6),
only the Roper resonance has a nonzero isoscalar longitudinal form factor.
All other three resonances cannot be excited by isoscalar longitudinal
virtual photons. The \underline{isoscalar} longitudinal amplitudes of
$S_{11}(1535)$ and $D_{13}(1520)$ vanish because of spin-flavor symmetry,
while both isoscalar and isovector longitudinal couplings of
$S_{11}(1650), D_{15}(1675)$ and $D_{13}(1700)$ vanish identically. This
behavior of the isoscalar form factors is essential for the correct
description of the existing experimental data on the $t-$dependence of
$T_{20}$ for the process $d+p\rightarrow d+X$.

The longitudinal isoscalar ($A_{S} = A_{\ell}^p + A_{\ell}^n$)
form factor of the Roper resonance in the
collective-string model of \cite{Bi94,Bi96} is,
apart from an overall constant,
\begin{equation}
A_S =\frac{4ka(1-2k^2a^2)}{(1+k^2a^2)^4} ~.
\end{equation}
This form factor has a  zero  at $ k^2=\displaystyle\frac{1}{2 {a^2}}$.
This is also the zero of $r$ seen in Fig. 2. The location of the zero depends 
on the value of $a$, which in turn
characterizes the size of the string configuration. An accurate
determination of the location of the zero for the transition form factor from
the
$t-$dependence of $T_{20}$ could therefore provide an independent
measure of the transition radius of the Roper resonance.
In particular, it could shed some light on the question whether or not
hadrons swell with increasing excitation energy. The latter can be studied
within the collective-string model \cite{Bi94,Bi96} by introducing the
stretchability of the string $\xi$ via the ansatz
\begin{equation}
a=a_0\left ( 1+\xi \displaystyle\frac{W-M}{M}\right )
\end{equation}
with $a_0=0.232$ fm.

The calculations reported below are performed in
the Breit frame for which
\begin{equation}
{k^2}=-t+\displaystyle\frac{(W^2-M^2)^2}{2(W^2+M^2)-t}.
\end{equation}
From Eq. (9), one can see that $k$ depends not only on $t$, but on $W$ too.
\section{Results and discussion}

In Fig. 3 we report the theoretical predictions, using Eq. (3), together with
the existing experimental data. In such an approximation, $T_{20}$ is a
universal function of $t$ only, without
any dependence on the initial deuteron momentum.
The experimental values of $T_{20}$ for
$p(\vec d,d)X$ \cite{lns250,Az96}, for different momenta of the incident
beam are shown as open symbols. These data show a scaling as a function of
$t$, with a small dependence on the incident momentum
in the interval 3.7--9 GeV/c. On the same plot the data for the elastic
scattering process $e^-+d\rightarrow e^-+d$ \cite{mg} are shown (filled stars).
All these data show a very similar behavior: negative values, with a minimum
in the region $|t|\simeq  0.35~GeV^2$ and an increase towards zero at
larger  $|t|$. The lines are the
result of the $\omega$-exchange model calculation for the $d+p\rightarrow d+X$
process. The dashed-dotted line correspond to $r = 0$, i.e. to
$T_{20}=-1/\sqrt{8}$ as mentioned previously.
Calculations based on the algebraic collective model \cite{Bi94,Bi96}
are shown for the case when only the Roper resonance is considered
(dotted line) and for the case when all the four resonances (6)
are considered (solid line). The required deuteron form factors,
$G_c,~G_q,~G_m$, were taken from \cite{chu} (calculated in
a relativistic impulse approximation) and they reproduce well the
$T_{20}-$data for $ed$ elastic scattering \cite{mg}.
When $r\gg 0$ or if the contribution of the deuteron magnetic form factor
$V_1$ is neglected, then $T_{20}$ does not depend on the ratio $r$, and
coincides with $t_{20}$ for the elastic $ed$-scattering (with the same
approximation).

From Fig. 3 it appears that the $t-$behavior of $T_{20}$ is very sensitive to
the value of $r$, at relatively small $r$, $r\leq 0.5$.
The values of $r$, predicted by the collective model \cite{Bi94,Bi96}
give a good description of the data, when taking into account the
contribution of all four resonances (6). These data, in any case, exclude a 
very small value of
$r,~r \ll 0.1$ as well as very large values of $r$.
Such sensitivity of $T_{20}$ for
$d+p\rightarrow d+X$ to the ratio of the corresponding isoscalar form
factors of the
$N^*$-excitation clearly indicates the presence of the Roper
resonance in this process.
Such an indication was hardly found in the differential cross section for
inclusive scattering with unpolarized particles \cite{Mo92}.

In the framework of the $\omega$-exchange mechanism,
electromagnetic isovector components of the $N\rightarrow N^*$ transition
cannot contribute. This is important as the other resonances in
this mass region, $S_{11}(1535),D_{13}(1520)~\mbox{ and} ~S_{11}(1650)$
are essentially isovector in the collective model \cite{Bi94,Bi96}
(as well as in other quark models), so the
isoscalar longitudinal form factors for $N\rightarrow N^*$ are identically zero
for any value of $t$. The ratio $r$ contains (in the numerator) the
contribution of only the Roper resonance. It
is this specific property of the Roper resonance (combined with the
$t-$dependence of the deuteron form factors) that induces the specific
$t-$behavior of the isoscalar ratio $r$ and of the analyzing power
$T_{20}$ as shown in Figs. 2 and 3, repectively.

In this connection, we mention that the $\omega$-exchange model predicts
the general features of the polarization observables. For example,
the crossing of all the theoretical curves for $T_{20}$ at two points,
 is
determined by the relative value of the deuteron
electromagnetic form factors. For any model of $r$ we will have
$T_{20}\leq -1/\sqrt{8}$ in the region $2\leq q\leq 5$~fm$^{-1}$.
Future data from Jefferson Lab \cite{Ko94}, concerning $T_{20}$ in
$e^-+d\rightarrow e^-+d$
will help in defining the exact position of these points. For $q\leq
6$~fm$^{-1}$, $T_{20}$ cannot be positive.

Using the generalized formula, Eq. (4), the $t-$behavior of the ratio
$r(t,W)$ depends on the initial deuteron momentum.
From Fig. 4 one can see that, in the interval 3.7--9 GeV/c this
dependence is not so large, in agreement with experimental data.
This is also true for the momentum dependence of $T_{20}$, (Fig. 5). The
agreement
between theoretical predictions and
experimental data is generally good, at least for
$q \leq$ 2 fm$^{-1}$.

Of course, this model for $d+p\rightarrow d+p$ can be improved, taking into
account
for example, other
meson exchanges,
or the effects of the strong interaction in initial and final states.
However these
corrections are strongly model- and parameter- dependent; the existing
experimental data are not
good enough to constrain the additional parameters which have to be added. In
this case we
lose the predictive power of our "parameter free" model. The successful
description of the
polarization observable $T_{20}$ can be considered as a strong  indication that
the $\omega-$
exchange is the main mechanism for the considered process.

We analyzed also the  sensitivity to a possible stretching mechanism
\cite{Bi96}, leading to the swelling of hadrons with increasing
excitation energy.
We use the parameterization of Eq. (8) for the scale parameter $a$,
with $\xi=0.5$ and $\xi=1$ (the last value is consistent with the
analysis of the experimental mass spectra, Regge trajectories).
The results are reported in Figs. 6 and 8 for $r(t,W)$ and
Figs. 7 and 9 for
$T_{20}$, for the different values of initial deuteron momentum.
The behavior of $r(t,W)$ is seen to be very sensitive to $\xi$. Introduction of
swelling
gives a more negative
slope to $T_{20}$ in better agreement with experiment although the position
of the  minimum  at $p_d$= 3.7 and 9 GeV/c is still shifted
to higher $q$ values compared to that measured by the data.

Similar results can be obtained for other polarization observables.
In Fig. 10 we show the $t-$dependence of the vector polarization transfer
coefficient, $K^{y'}_y$, from the initial to the scattered deuteron.
It is characterized by a strong sensitivity to the ratio $r$
for $q\geq 3$~fm$^{-1}$. This observable is especially interesting in this
region, because $T_{20}$ vanishes around $q\simeq 5$~fm$^{-1}$ (for any value 
of $r$). Our calculations
predict quite a large absolute value of this
observable and a strong dependence on the variable $t$.

Let us note in this connection, that, all T-even polarization observables
are nonzero and large in absolute value. This is an intrinsic property of
$\omega$-exchange. In contrast, all T-odd polarization effects cancel,
because we neglected the effects of strong interaction in
initial and final states. However, for collinear kinematics, all spin-one T-odd
polarization observables must be zero, in any model. The most simple
T-odd polarization observable, which exists in the general case for the
collinear kinematics, corresponds to the
correlation coefficient $\vec u\cdot\vec P\times\vec Q$,
$Q_a=Q_{ab}u_b$, where $\vec u$ is
the unit vector along the initial 3-momentum, $\vec P$ is the proton
polarization and $Q_{ab}$
is the deuteron tensor polarization. A measurement of these observables will
give a direct information on the presence and intensity of the final or
initial strong interaction. The tensor analyzing power $T_{20}$, being a
T-even observable, is less sensitive to the  effects
of strong interaction in initial and final states, as these effects induce
`quadratic corrections' to any T-even polarization observable.
We can schematically write: $P^{(+)}$=$P_0^{(+)}(1+\delta^2)$,
where $P^{(+)}$ is any T-even polarization observable and
$\delta$ represents the correction due to rescattering effects.
Any T-odd polarization
observable, $P^{(-)}$, is directly proportional to $\delta$.

\section{ Conclusions}
We have shown the sensitivity of the tensor analyzing power in the process
$d+p\rightarrow
d+X$ (for forward deuteron scattering) to the relative value of the cross
sections for the
absorption of virtual isoscalar photon ($\gamma^*+N\rightarrow N^*$) with
longitudinal and transversal polarizations.
The main point is that only the Roper resonance excitation is characterized 
by a nonzero
isoscalar longitudinal form factor, whose $t-$behavior drives the
$t-$dependence of the
tensor analyzing power. The other resonances, lying in this mass region, 
such as $S_{11}(1535)$,
$D_{13}(1520)$  and $S_{11}(1650)$ are characterized, due to
the specific quark structure, by a pure isovector nature of longitudinal 
virtual photons
absorbed by nucleons. Without excitation of the Roper resonance,  $r=0$, and
the value for
$T_{20}$ becomes $t-$independent: $T_{20}=-1/2\sqrt{2}$, in evident
disagreement with existing data. The specific behavior of $r$ obtained
in the framework of the collective model \cite{Bi94,Bi96}
for the Roper resonance, and in the presence of other resonances, with a
definite
isotopic structure, are very important for obtaining a good description
of the data.
In the framework of the $\omega-$exchange mechanism, the $t-$behavior of
$T_{20}$ for the
reaction $p(\vec d, d')X$, is affected, on one side, from the specific $t-$
dependence of all
three deuteron electromagnetic form factors, and on the other side, from the
values of $r$
when all four overlapping resonances
$P_{11}(1440),~S_{11}(1535),D_{13}(1520)~\mbox{ and}
~S_{11}(1650)$ contribute. The main property is the role of the Roper resonance
excitation,
with nonzero longitudinal isoscalar form factor.

Also to be noticed is that the sensitivity of $T_{20}$ to $r$ is especially
large for
$r$ in the interval $r \leq 0.5$, which is in agreement with the data
on inclusive $eN$-scattering at these values of excitation energy. Our model
predicts a small momentum dependence for $T_{20}$, consistent with the
experimental data. The results are also stable with
respect to the parameter $\xi$, the stretchability of the nucleon string,
in the range $\xi=0.5-1$.

The study of polarization observables in the reaction $d+p\rightarrow d+X$
can be considered as an additional method to measure the
$t-$dependence of the ratio $r$. Moreover, the process
$p(d,d')$X is sensitive to the isoscalar contributions to $r$. The vector
polarization transfer coefficient $K^{y'}_y$ is also sensitive to this ratio,
in a different region of $t$.

We can consider the existing data about $T_{20}$ not only as an evidence for
the Roper resonance excitation, but also as a tool to study properties
of isoscalar form factors for the excitation of $N^*$ resonances,
complementary to the inelastic electron-nucleon scattering,
$e^-+N\rightarrow e^-+N^*$.
The possibility to unify in a common picture such different processes, as
$e^-+d\rightarrow
e^-+d$ and $e^-+N\rightarrow e^-+N^*$, from one side, and a  hadronic
process as $d+p\rightarrow d+X$ from another side, suggests a new
perspective to study nucleon structure
through electromagnetic and hadron excitation of nucleonic resonances.

At large energies of colliding particles, instead of $\omega-$exchange
it is necessary to consider Pomeron (${\cal P}$) exchange, i.e. the mechanism 
of diffractive excitation of $N^*$, in $dp$-forward scattering.
Concerning polarization phenomena, the properties of $\omega$ and
${\cal P}$- are similar. But in the general case,
the vertexes of ${\cal P}NN^*$ and ${\cal P}dd$-interactions are different
from $\omega NN^*$ and $\omega dd$-vertexes:
only in the framework of the hypothesis about {\it Pomeron-photon analogy},
the prediction of these models coincide, at least concerning polarization
phenomena.

\section{ACKNOWLEDGMENTS}

This work is supported in part
by DGAPA-UNAM under project IN101997 (R.B.), by grant No. 94-00059 from the
United States-Israel Binational Science Foundation (BSF),
Jerusalem, Israel (A.L.) and by D.O.E. Grant DE-FG02-91ER40608 (F.I.).


%
%
\newpage
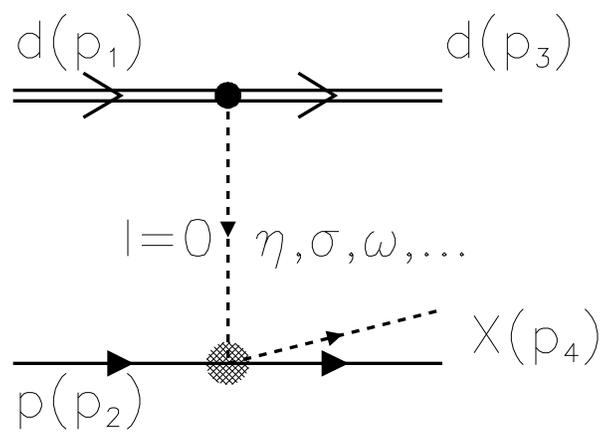
\begin{figure}
 \caption{ $t-$channel meson exchange for $d+p\rightarrow d+X$}.
\end{figure}
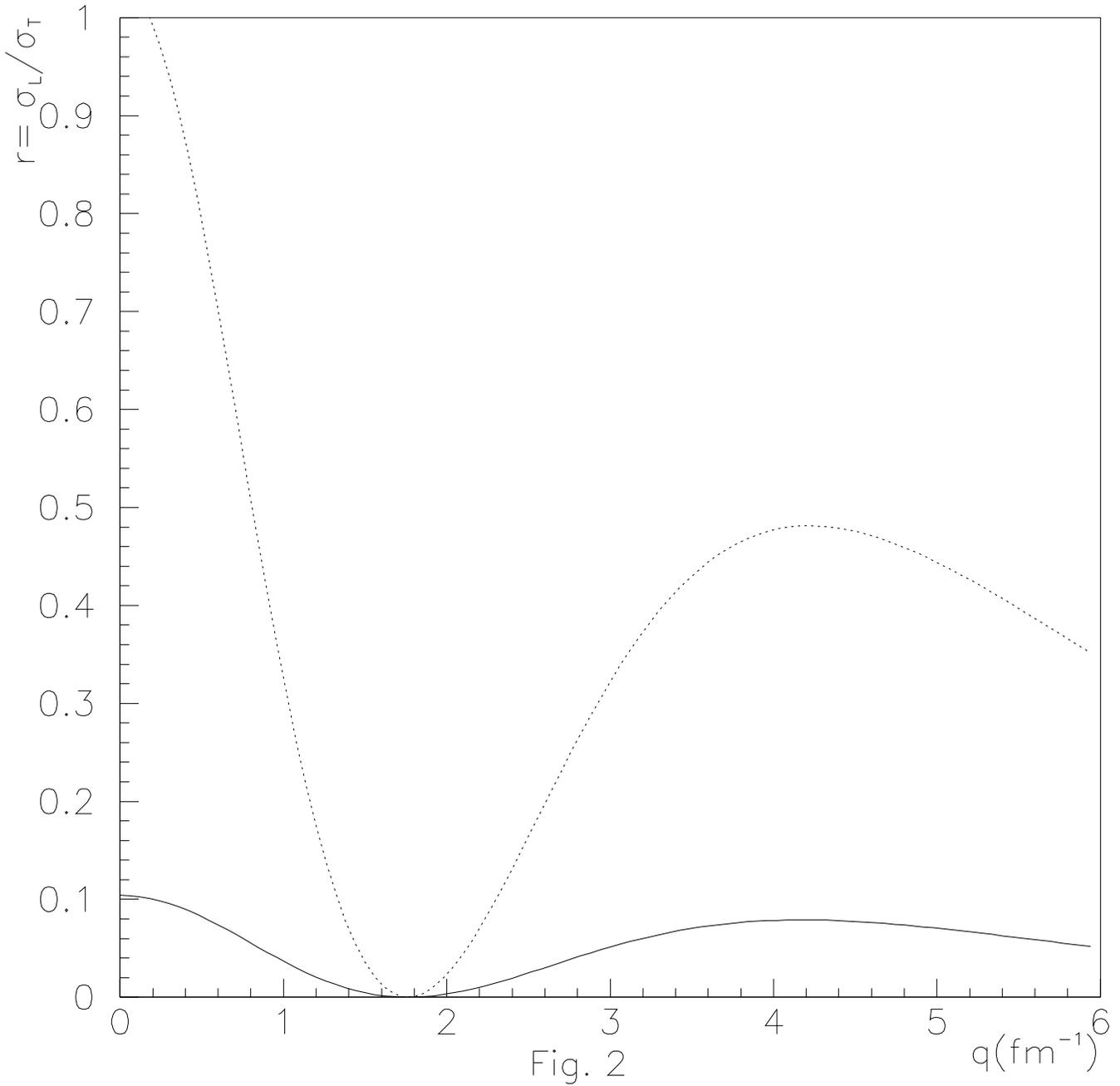
\begin{figure}
\caption{The ratio $r$ for the case when only the
Roper  excitation is considered (dotted line) and for the case when
all four resonances  (6) are considered, (solid line), from Eq. (3).}
\end{figure}
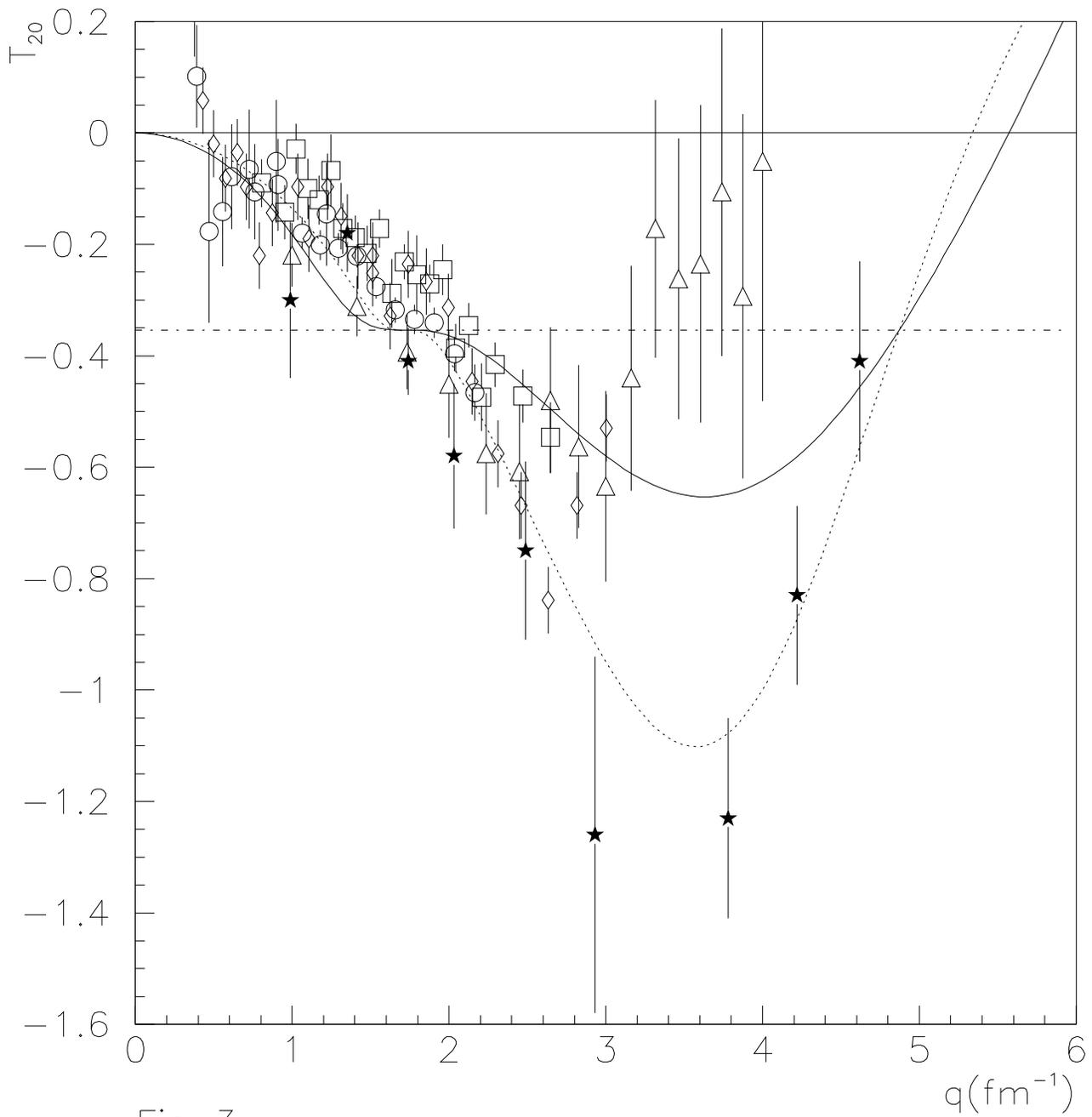
\begin{figure}
\caption{Experimental data for $T_{20}$ for
$e^-+d\rightarrow e^-+d$ elastic scattering (filled stars) [7]  
and $ d+p\rightarrow d +X$ at incident momenta of 3.75 GeV/c
 (open diamonds) [5],  
5.5 GeV/c (open circles),
4.5 GeV/c  (open squares),  9 GeV/c
 (open triangles) [6].  
Prediction of the $\omega-$exchange model for  $r=0$
(dashed-dotted line).
Calculations with $r$ using collective form factors (Tables I-II)
are shown for the case when only the Roper resonance is considered
(dotted line) and for the case when all the four
resonances  (6) are considered (solid line).}
\end{figure}
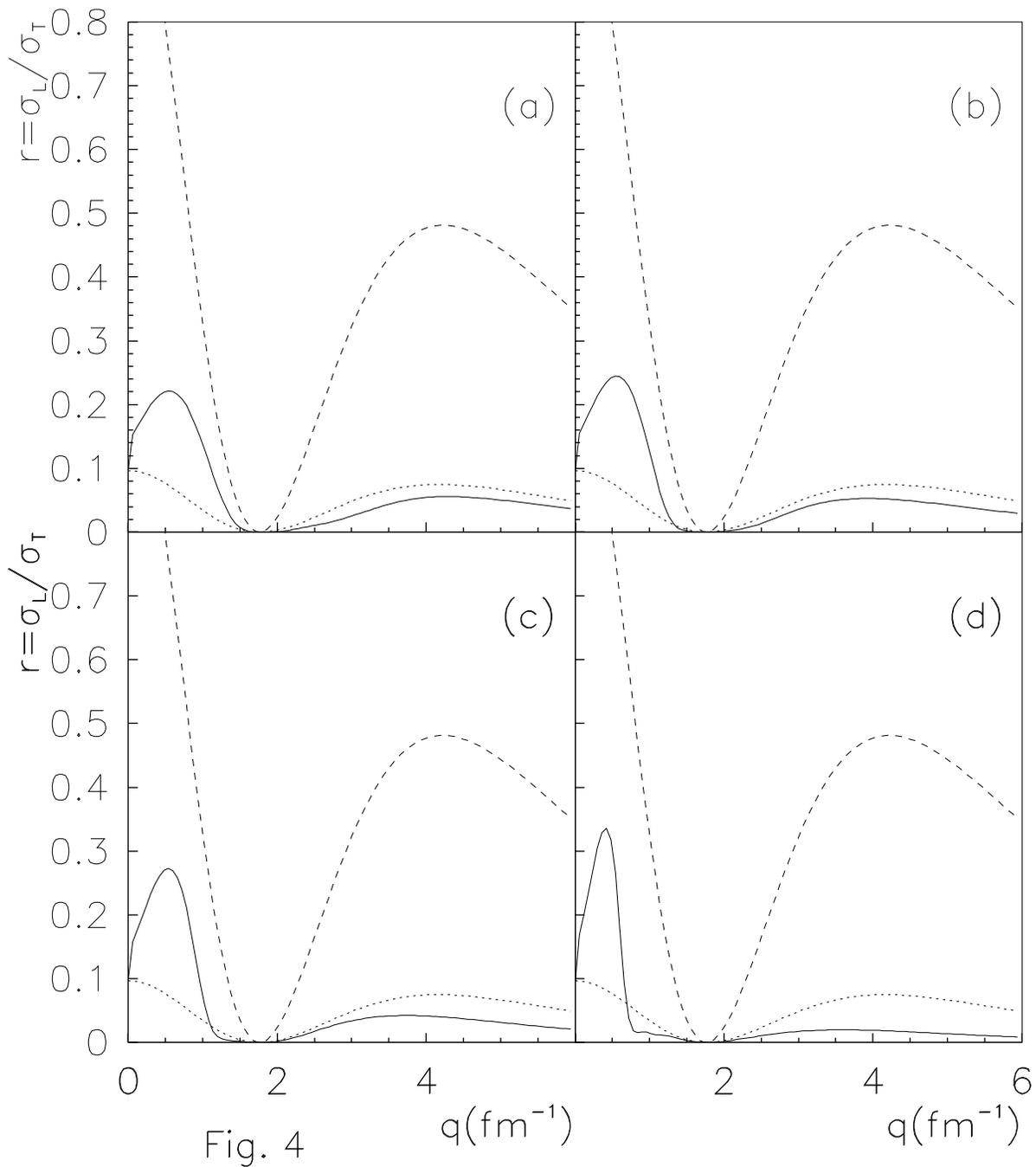
\begin{figure}
\caption{ $q-$ dependence of $r=\sigma_L/\sigma_T$ for excitation of
only the Roper resonance (dashed line),
for excitation of all resonances  (6) with (solid line)
and without (dotted line) width effects, for different deuteron momenta
$p_d$: (a) $p_d$=3.7 GeV/c, (b) $p_d$=4.5 GeV/c, (c) $p_d$=5.5 GeV/c,
(d) $p_d$=9 GeV/c.}
\end{figure}

\begin{figure}
\caption{ $q-$dependence of $T_{20}$ for different deuteron
momenta $p_d$: (a) $p_d$=3.7 GeV/c, (b) $p_d$=4.5 GeV/c, (c) $p_d$=5.5 GeV/c,
(d) $p_d$=9 GeV/c.}
\end{figure}
\begin{figure}
 \caption{ $q-$dependence of $r=\sigma_L/\sigma_T$ for $\xi=0.5$ 
(same notations as in Fig. 4).}
\end{figure}
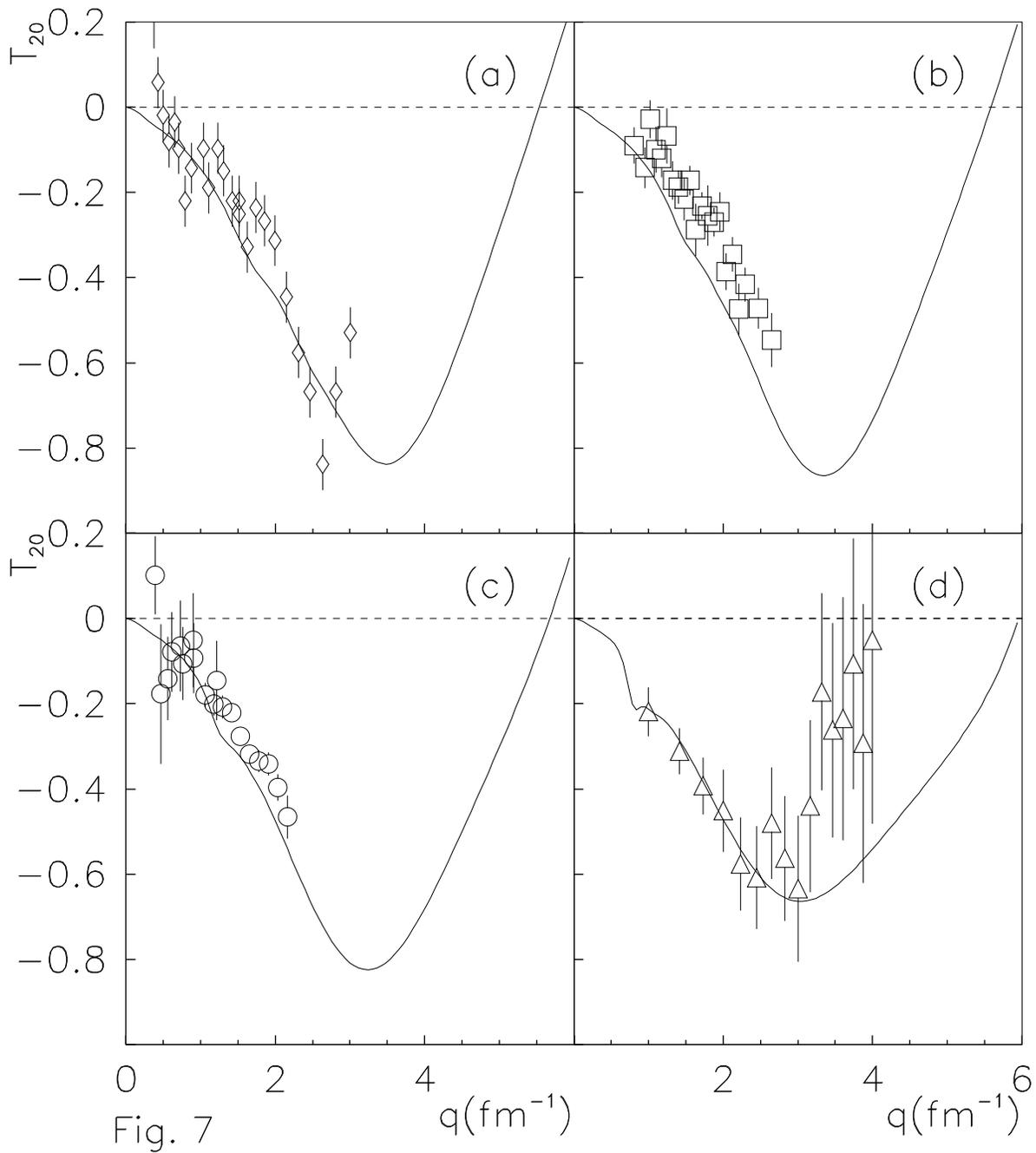
\begin{figure}
 \caption{ $q-$dependence  of $T_{20}$ for $\xi=0.5$ (same notations
as in Fig. 5).}
\end{figure}
\begin{figure}
 \caption{ $q-$ dependence of $r=\sigma_L/\sigma_T$ for $\xi=1$ (same notations
as in
Fig. 4).}
\end{figure}
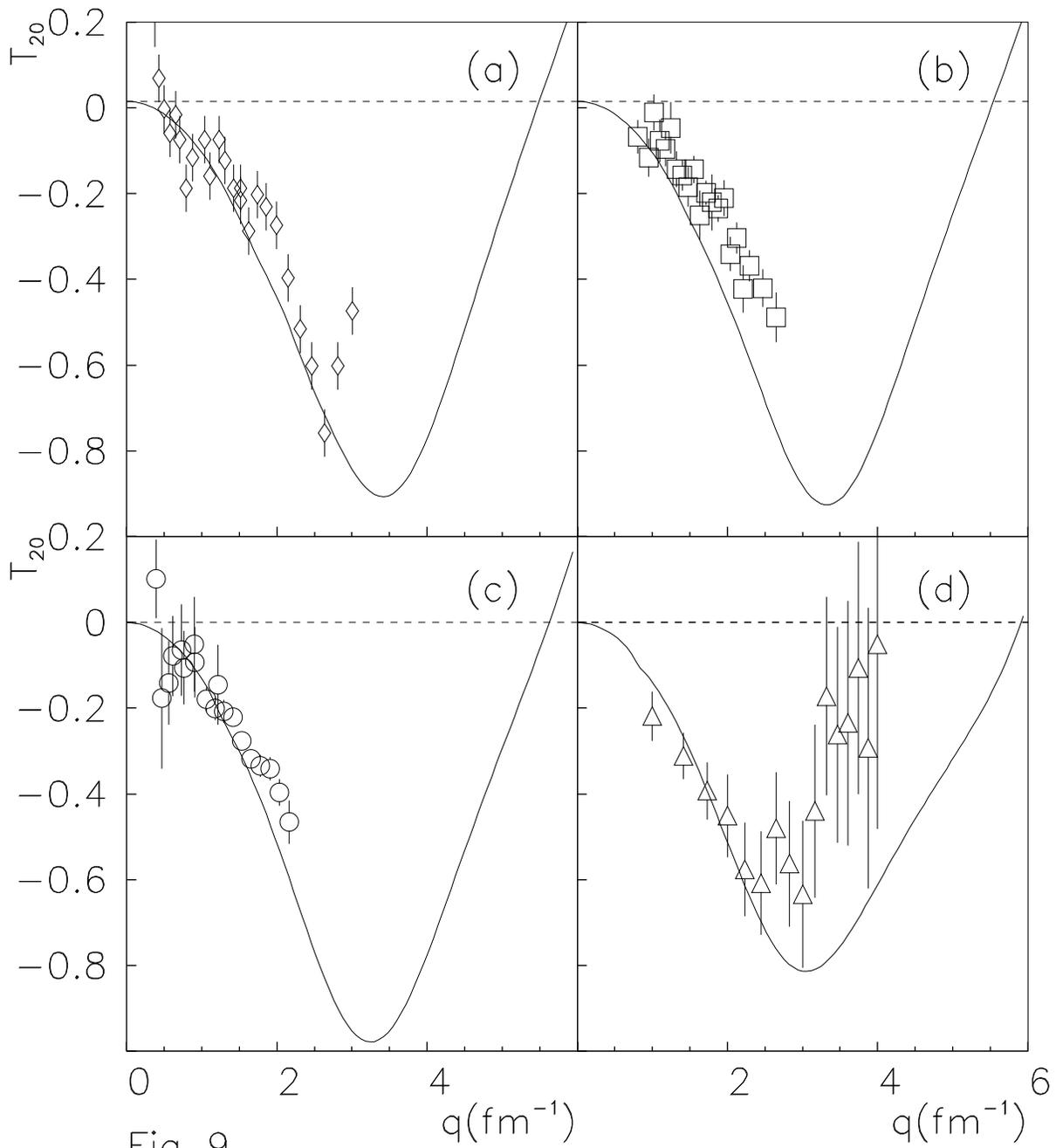
\begin{figure}
 \caption{ $q-$ dependence of $T_{20}$  for $\xi=1$ (same notations
 as
in Fig. 5).}
\end{figure}
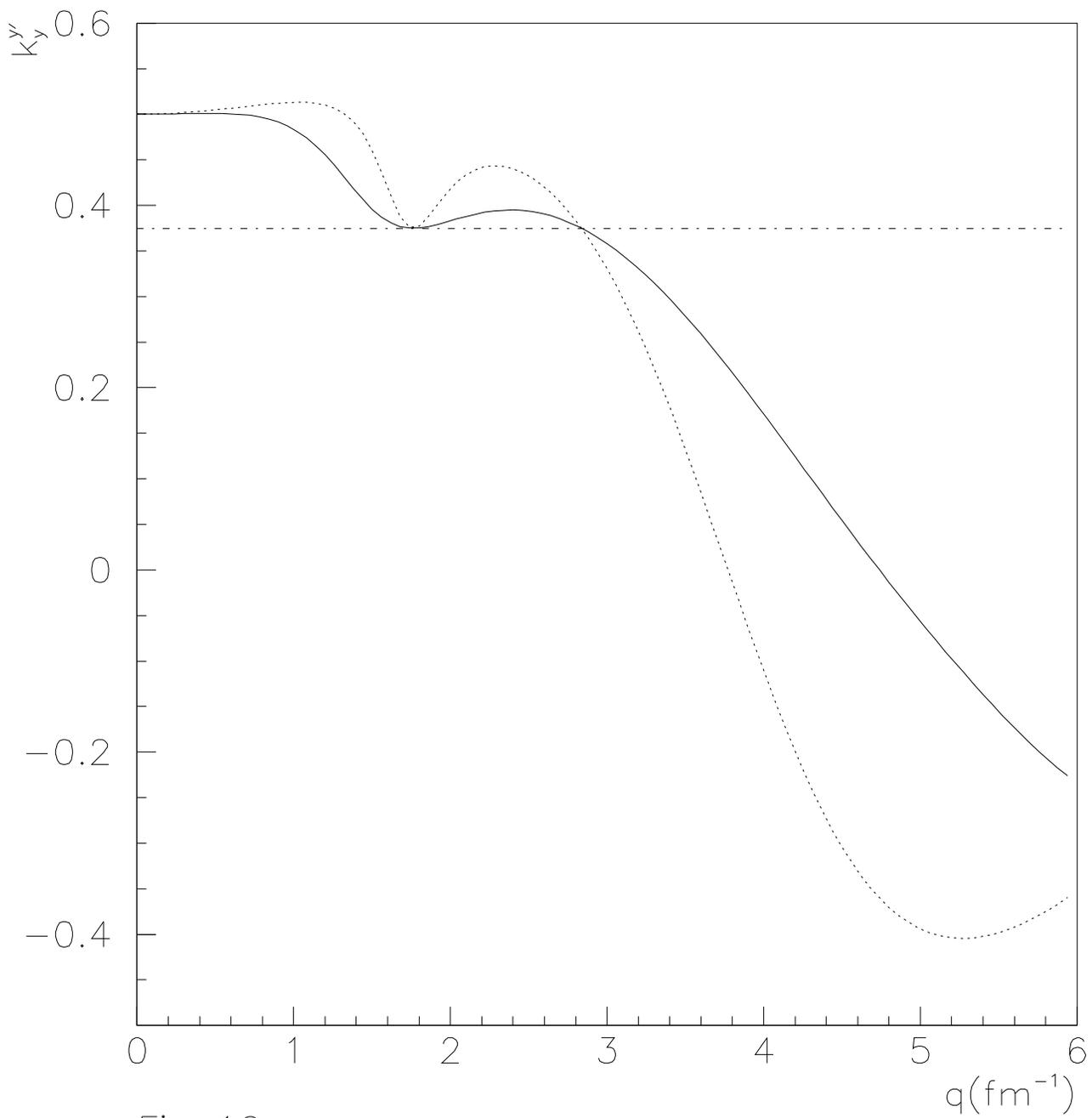
\begin{figure}
 \caption{ Vector polarization transfer coefficient
$K_y^{y'}$ as a function of
$q$, for $ d+p\rightarrow d +X$. Prediction of the $\omega-$exchange
model for $r=0$ (dashed-dotted line).
Calculations using the collective form factors (Tables I-II) are
shown for the case when only the Roper resonance is considered
(dotted line) and for the case when all the four resonances  (6)
are considered (solid line).}
\end{figure}

\clearpage
\renewcommand{\baselinestretch}{1.35}

\begin{table}
\centering
\caption[Transverse helicity amplitudes]{\small
Transverse proton ($A^{p}_{\nu}$) and neutron ($A^{n}_{\nu}$)
helicity amplitudes ($\nu=1/2,3/2$) of nucleon resonances below 2 GeV in the
collective model \cite{Bi94,Bi96}.
$Z(x) = -\frac{1}{(1+x^2)^2} + \frac{3}{2x^3}H(x)$ with
$H(x) = \arctan x - {x\over (1+x^2)}$.
$\chi_1$ and $\chi_2$ are parameters \cite{Bi96}.
\normalsize} \label{tamp_p}
\vspace{15pt}
\begin{tabular}{cccc}
\hline
& & & \\
Resonance & $\nu$ & $A^{p}_{\nu}$ & $A^{n}_{\nu}$ \\
& & & \\
\hline
& & & \\
$N(1440)P_{11}$ & $1/2$ & $-2 \chi_1 \sqrt{\pi\over k_0}\mu k \,
\frac{2k^2a^2}{(1+k^2a^2)^3}$ & ${4\over 3} \chi_1 \sqrt{\pi\over k_0}\mu k \,
\frac{2k^2a^2}{(1+k^2a^2)^3}$ \\
& & & \\
$N(1520)D_{13}$ & $1/2$ & $2i\sqrt{\pi\over k_0}\mu\frac{1}{(1+k^2a^2)^2}
\Bigl [{mk_0a\over g} - k^2a\Bigr ]$
& $-2i\sqrt{\pi\over k_0}\mu\frac{1}{(1+k^2a^2)^2}
\Bigl [{mk_0a\over g} - {1\over 3}k^2a\Bigr ]$ \\
& $3/2$ & $2i\sqrt{3}\sqrt{\pi\over k_0}\mu
\frac{1}{(1+k^2a^2)^2}{mk_0a\over g}$
& $-2i\sqrt{3}\sqrt{\pi\over k_0}\mu
\frac{1}{(1+k^2a^2)^2}{mk_0a\over g}$ \\
& & & \\
$N(1535)S_{11}$ & $1/2$ & $i\sqrt{2}\sqrt{\pi\over k_0}\mu
\frac{1}{(1+k^2a^2)^2} \Bigl [2{mk_0a\over g} + k^2a\Bigr ]$
& $-i\sqrt{2}\sqrt{\pi\over k_0}\mu\frac{1}{(1+k^2a^2)^2}
\Bigl [2{mk_0a\over g} + {1\over 3}k^2a\Bigr ]$ \\
& & & \\
$N(1650)S_{11}$ & $1/2$ & 0 & $i{\sqrt{2}\over 3}
\sqrt{\pi\over k_0}\mu k \, \frac{ka}{(1+k^2a^2)^2}$ \\
& & & \\
$N(1675)D_{15}$ & $1/2$ & 0 & $-i{\sqrt{2\over 5}} \sqrt{\pi\over k_0}
\mu k \, \frac{ka}{(1+k^2a^2)^2}$ \\
& $3/2$ & 0 & $-i{2\over\sqrt{5}}\sqrt{\pi\over k_0}\mu k\,
\frac{ka}{(1+k^2a^2)^2}$ \\
& & & \\
$N(1680)F_{15}$ & $1/2$ & $-\sqrt{3}\sqrt{\pi\over k_0}\mu
\Bigl [ 2{mk_0a\over g} - k^2a\Bigr] {1\over ka} Z(ka)$
& $-{2\over\sqrt{3}}\sqrt{\pi\over k_0}\mu k Z(ka)$ \\
& $3/2$ & $-2\sqrt{6} \sqrt{\pi\over k_0}\mu \,
{mk_0a\over g} \, {1\over ka} Z(ka)$ & 0 \\
& & & \\
$N(1700)D_{13}$ & $1/2$ & 0 & $i{1\over 3}{\sqrt{2\over 5}}
\sqrt{\pi\over k_0}\mu k\,\frac{ka}{(1+k^2a^2)^2}$ \\
& $3/2$ & 0 & $ i{\sqrt{6\over 5}}\sqrt{\pi\over k_0}\mu k\,
\frac{ka}{(1+k^2a^2)^2}$ \\
& & & \\
$N(1710)P_{11}$ & $1/2$ & $\sqrt{2}\chi_{2}\sqrt{\pi\over k_0} \mu
\frac{2k^2a^2}{(1+k^2a^2)^3}$ & $-{\sqrt{2}\over 3}\chi_{2}
\sqrt{\pi\over k_0}\mu \frac{2k^2a^2}{(1+k^2a^2)^3}$  \\
& & & \\
$N(1720)P_{13}$ & $1/2$ & $-\sqrt{2}\sqrt{\pi\over k_0}\mu
\Bigl [ 3{mk_0a\over g} + k^2a\Bigr] {1\over ka} Z(ka)$
& ${2\sqrt{2}\over 3}\sqrt{\pi\over k_0}\mu k Z(ka)$ \\
& $3/2$ & $\sqrt{6} \sqrt{\pi\over k_0}\mu \,
{mk_0a\over g} \, {1\over ka} Z(ka)$ & 0 \\
& & & \\
\hline
\end{tabular}
\end{table}

\newpage

\begin{table}
\centering
\caption[Longitudinal helicity amplitudes]{\small
 Longitudinal  proton ($A^{p}_{\ell}$) and
neutron ($A^{n}_{\ell}$) helicity
amplitudes of nucleon resonances below 2 GeV in the collective model
\cite{Bi94,Bi96}. Notation as in Table~\ref{tamp_p}.
\normalsize} \label{tamp_nl}
\vspace{15pt}
\begin{tabular}{cccc}
\hline
& & &  \\
Resonance & State & $A^{p}_{\ell}$ & $A^{n}_{\ell}$ \\
& & & \\
\hline
& & & \\
$N(1440)P_{11}$ & $^{2}8_{1/2}[56,0^+]_{(1,0);0}$
& $2 \chi_1 \sqrt{\pi\over k_0}\mu {mk_0a\over g}
{4ka ( 1 -2k^2a^2)\over {(1+k^2a^2)^4}}$
& $0$ \\
& & & \\
$N(1520)D_{13}$ & $^{2}8_{3/2}[70,1^-]_{(0,0);1}$
& $2 i \sqrt{\pi\over k_0}\mu {mk_0a\over g}
{1 -3k^2a^2 \over {(1+k^2a^2)^3}}$
& $A^{n}_{\ell} = -A^{p}_{\ell}$ \\
& & & \\
$N(1535)S_{11}$ & $^{2}8_{1/2}[70,1^-]_{(0,0);1}$
& $-i\sqrt{2} \sqrt{\pi\over k_0}\mu {mk_0a\over g}
{1 -3k^2a^2 \over {(1+k^2a^2)^3}}$
& $A^{n}_{\ell} = -A^{p}_{\ell}$ \\
& & & \\
$N(1650)S_{11}$ & $^{4}8_{1/2}[70,1^-]_{(0,0);1}$
& $0$ & $0$ \\
& & & \\
$N(1675)D_{15}$ & $^{4}8_{5/2}[70,1^-]_{(0,0);1}$
& $0$ & $0$ \\
& & & \\
$N(1680)F_{15}$ & $^{2}8_{5/2}[56,2^+]_{(0,0);0}$
& $-\sqrt{3}\sqrt{\pi\over k_0}\mu {mk_0a\over g}
\Bigl [ {4ka\over (1+k^2a^2)^3} -{3\over ka} Z(ka) \Bigr ]$
& $0$ \\
& & & \\
$N(1700)D_{13}$ & $^{4}8_{3/2}[70,1^-]_{(0,0);1}$
& $0$ & $0$ \\
& & & \\
$N(1710)P_{11}$ & $^{2}8_{1/2}[70,0^+]_{(0,1);0}$
& $-\sqrt{2}\chi_{2}\sqrt{\pi\over k_0}\mu {mk_0a\over g}
{4ka( 1 -2k^2a^2)\over (1+k^2a^2)^4}$
& $A^{n}_{\ell} = -A^{p}_{\ell}$ \\
& & & \\
$N(1720)P_{13}$ & $^{2}8_{3/2}[56,2^+]_{(0,0);0}$
& $ \sqrt{2}\sqrt{\pi\over k_0}\mu {mk_0a\over g}
\Bigl [ {4ka\over (1+k^2a^2)^3} -{3\over ka} Z(ka) \Bigr ]$
& $0$ \\
& & & \\
\hline
\end{tabular}
\end{table}

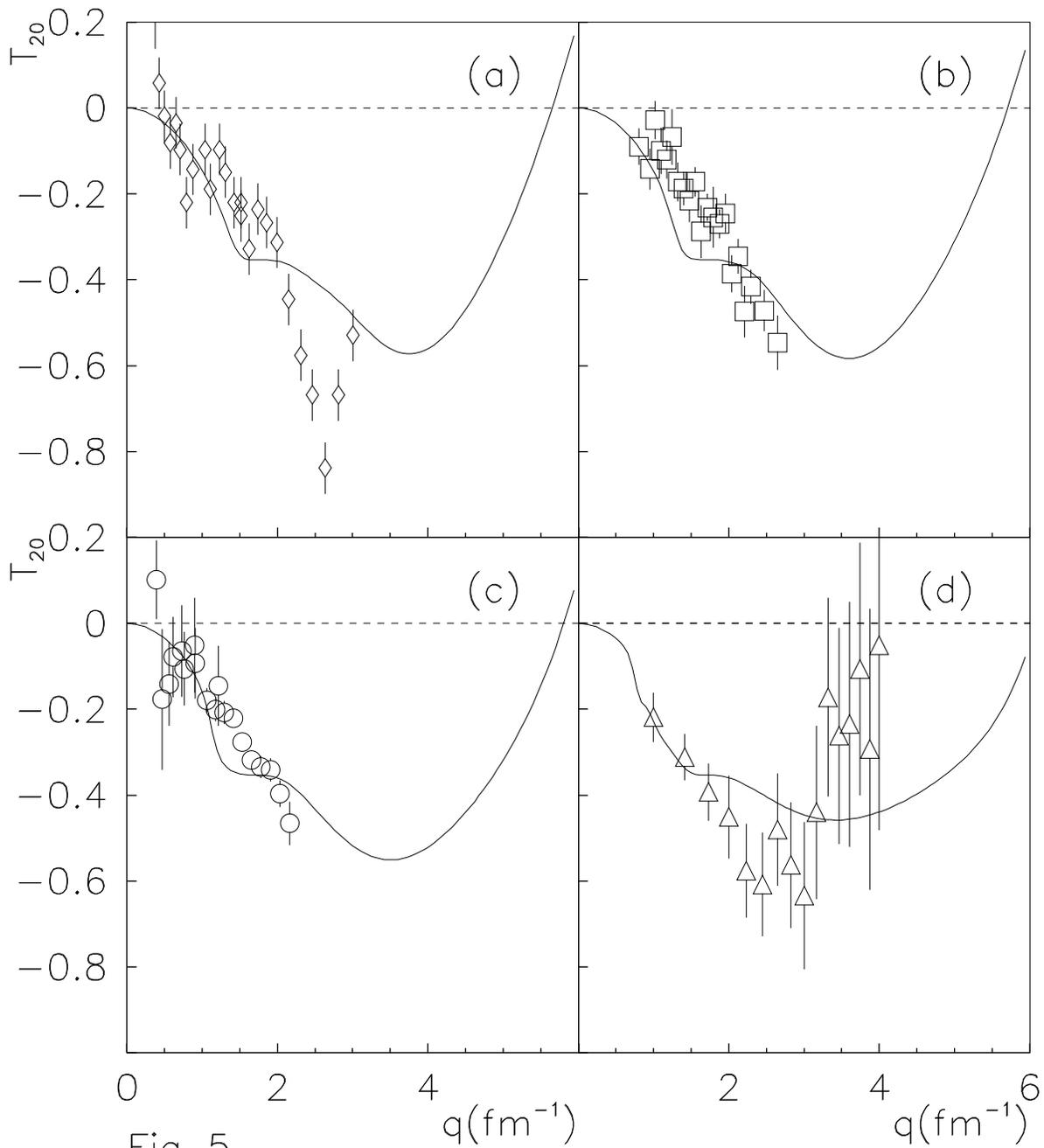
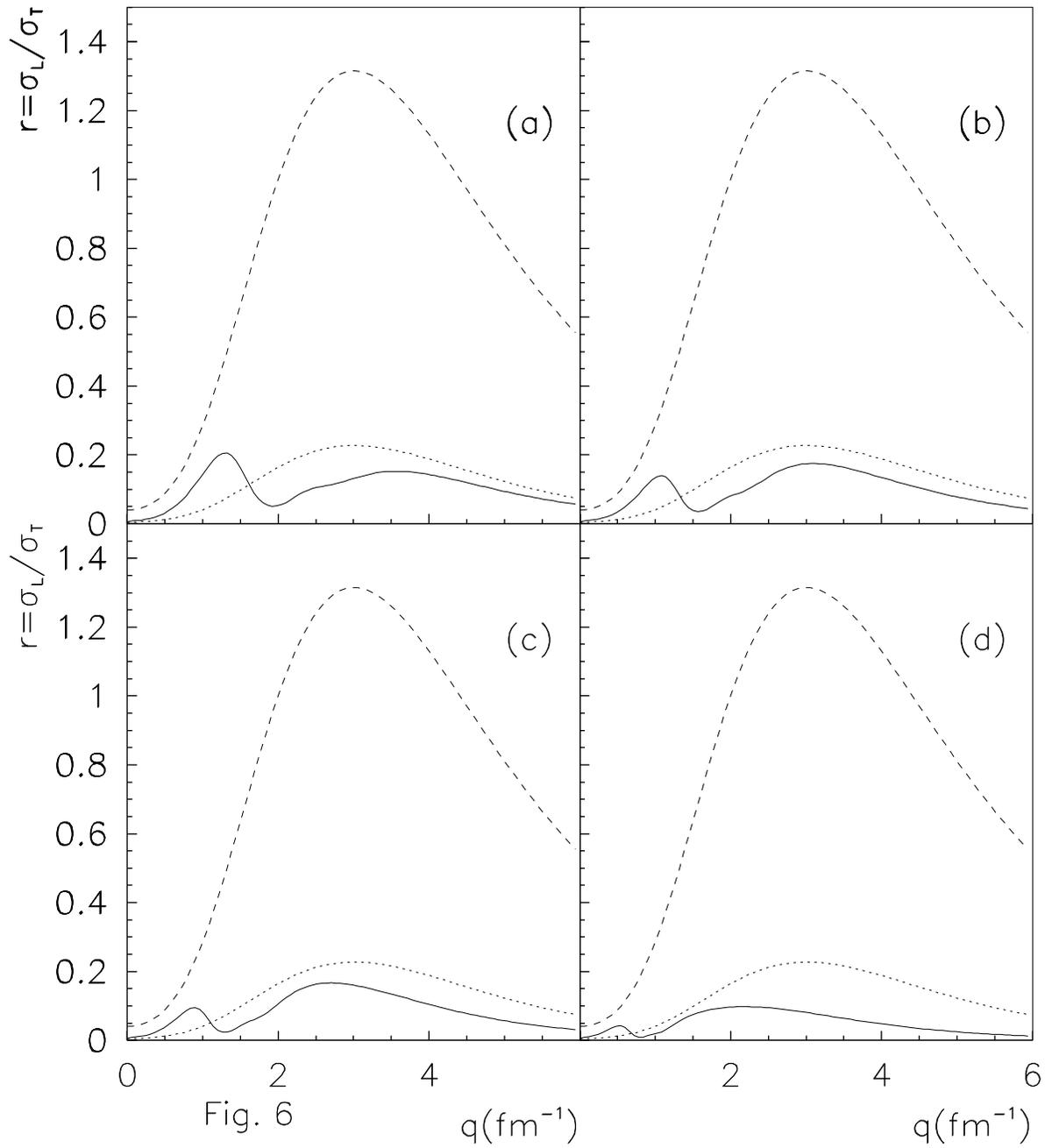
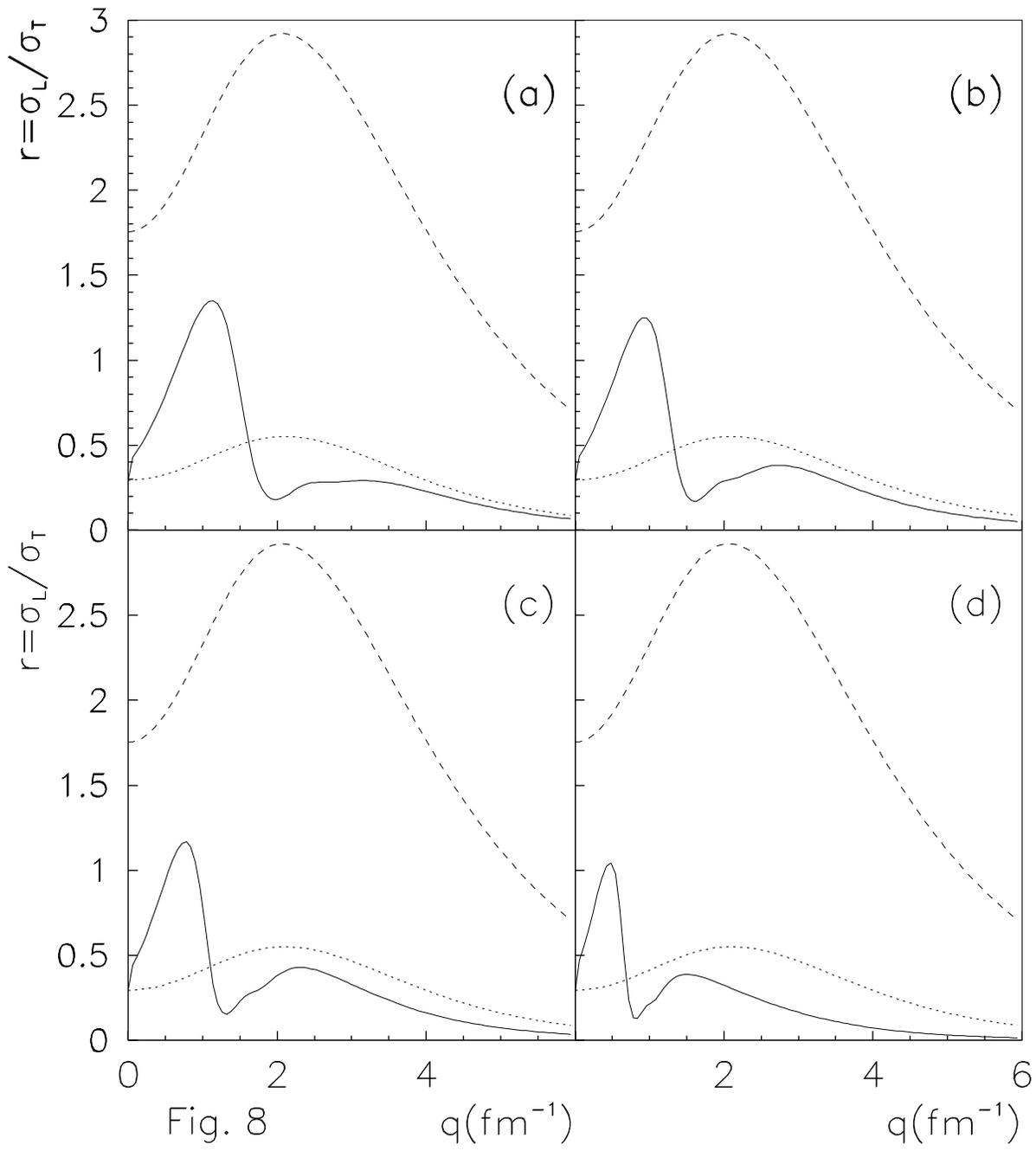
\end{document}